\documentclass[final,twocolumn]{revtex4}

\usepackage{graphicx}

\begin{document}





\title{Isospin equilibration in multi-fragmentation processes and dynamical correlations}

\author{M. Papa$^{a)}$ and G. Giuliani$^{b)}$ } 


\affiliation{a)Istituto Nazionale Fisica Nucleare-Sezione
di Catania, V. S.Sofia 64 95123 Catania Italy} 
\affiliation{b)Dipartimento di Fisica e Astronomia, Universit\'a di Catania V. S.Sofia 64 95123 Catania Italy}

\begin{abstract}
The asymptotic time derivative of the total dipole signal is
proposed as an useful observable to study Isospin equilibration
phenomenon in multi-fragmentation processes. The study proceeds
through the investigation of the $^{40}Cl+^{28}Si$ system at 40
MeV/nucleon by means of semiclassical microscopic many-body
calculations based on the CoMD-II model. In particular, the study
has been developed to describe charge/mass equilibration processes
involving the gas and liquid "phases" of the total system formed
during the early stage of a collision. Through the investigation of
dynamical many-body correlations, it is also shown how the proposed
observable is rather sensitive to different parameterizations of the
isospin dependent interaction.

\end{abstract}
\maketitle







\section{Introduction}
An interesting subject related to Heavy Ions Isospin physics
\cite{schr} is the process  leading to the equilibration of the
charge/mass ratio between the main partners of the reaction as well
described in Ref.\cite{pawel}. The so called "isospin diffusion"
phenomenon is the relevant mechanism acting between the reaction
partners in binary processes\cite{betty,baodif,dtoro,gal}. In
particular, in  the collision of the 124 and 112 Tin isotopes at 50
MeV/nucleon \cite{betty}, evidence of partial equilibrium in the
charge/mass ratios of the quasi-projectile and quasi-target has been
deduced through the study of the iso-scaling parameters related to
the isotopic distributions. In this case dynamical calculations
based on the Boltzmann-Uehling-Uhlenbeck model \cite{buu} show that
the degree of equilibration depends on the behavior of the symmetry
potential $U^{\tau}$ as a function of the density. The analysis of
the experimental data in this kind of studies is based on the linear
relation between the iso-scaling parameter and the relative neutron
excess $\beta$ of the emitting sources (typical in several
statistical models). It is also assumed that both quantities weakly
depend on the secondary statistical decay processes. In general, as
discussed in ref.\cite{betty1},
 this last condition can be affected by the fragment
excitation energies, (or temperatures) and by the distinctive
features  of the
 models used to simulate the secondary decay stage.
In this work we want to extend the study of the isospin
equilibration processes, looking at the whole system, by using the
following quantity
$\overrightarrow{V}(t)=\sum_{i=1}^{Z_{tot}}\overrightarrow{v}_{i}$.
The sum on the index $i$ is performed  on all the $Z_{tot}$ protons
of the system. $\overrightarrow{V}(t)$ corresponds, apart from the
elementary charge $e$, to the time derivative of the total dipole of
the system. The velocities $\overrightarrow{v_{i}}$ are computed in
the center of mass (c.m.) reference frame. Several studies were
based on this dynamical variable to describe pre-equilibrium
$\gamma$-ray emission (see Refs.\cite{asygdr,ca10mev,trasex1} and
references therein ). Various reasons suggest us to use the same
variable to also describe isospin equilibration processes.

 - (i) After the pre-equilibrium
 stage, starting from the time $t_{pre}$, when a second stage characterized by
 an average isotropic emission of the secondary sources (statistical equilibrium)
 takes place, the ensemble average of $\overrightarrow{V}(t)$
 satisfies the following relation:
 $\overline{\overrightarrow{V}}(t_{pre})=
 \overline{\overrightarrow{V}}(t>t_{pre})\equiv\overline{\overrightarrow{V}}$ \cite{ca10mev}.
The average value of this dynamical variable at $t_{pre}$
 is in fact invariant with respect to statistical processes
 and therefore the value of $\overline{\overrightarrow{V}}$ is determined
only by the complex dynamics which characterizes the early stage of
the collision, when fast changes of the average nuclear density are
expected. In particular, $\overline{\overrightarrow{V}}$
 can be expressed as a function of the charge $Z$, mass $A$, average multiplicity
 $\overline{m}_{Z,A}$ and the mean momentum
 $\langle \overrightarrow{P}\rangle_{Z,A}$ of the detected particles having charge $Z$
 and mass A in the generic event:
\begin{eqnarray}
\overline{\overrightarrow{V}}=\sum_{Z,A}\frac{Z}{A}\overline{m_{Z,A}}
\overline{\langle \overrightarrow{P} \rangle}_{Z,A}
C_{\langle \overrightarrow{P}\rangle}^{Z,A}\\
C_{\langle \overrightarrow{P}\rangle}^{Z,A}=
\frac{ \overline{m_{Z,A}\langle \overrightarrow{P}\rangle_{Z,A}}}
{\overline{\langle \overrightarrow{P}\rangle}_{Z,A}\overline{m}_{Z,A}}
\end{eqnarray}
$C_{\langle \overrightarrow{P}\rangle}^{Z,A}$ is the correlation
function between the multiplicity and the mean momentum. This
correlation function plays a key role for the invariance property
and therefore requires for an  event by event analysis in which
many-body correlations can not be neglected.

For symmetry reasons, $\overline{\overrightarrow{V}}$ lies on the
reaction plane. It is directly linked with a  weighted mean of the
charge/mass ratio, as Eq.(1) suggests. It also takes into account
the average isospin flow direction through the momenta
$\overline{\langle \overrightarrow{P}\rangle}_{Z,A}$.

 -(ii) In the general case , we
find attractive the following decomposition:
$\overline{\overrightarrow{V}} = \overline{\overrightarrow{V}_{G}}+
\overline{\overrightarrow{V}_{L}}+\overline{\overrightarrow{V}_{GL}}$
where $\overline{\overrightarrow{V}_{G}}$ and
$\overline{\overrightarrow{V}_{L}}$ are the average dipolar signals
associated to the gas "phase" (light charged particles) and to the
"liquid" part \cite{indra}, corresponding to the motion of the
produced heavy fragments. The signal
$\overline{\overrightarrow{V}_{GL}}$ is instead associated to the
relative motion of the two "phases". By supposing, for simplicity,
that the gas "phase" is formed by neutrons and protons,
$\overline{\overrightarrow{V}}$ can be further decomposed as:
\begin{equation}
\overline{\overrightarrow{V}}=
\overline{\frac{A_{G}(1-\beta^{2}_{G})}{4}\overrightarrow{v}_{r}^{NP}}+
\overline{\frac{\mu_{G,L}(\beta_{L}-\beta_{G})}{2}\overrightarrow{v}_{cm,LG}}+\overline{\overrightarrow{V}_{r,L}}
\end{equation}
In the above expression the first term represents the contribution
related to the neutron-proton relative motion of the gas "phase"
expressed through the relative velocity
$\overrightarrow{v}_{r}^{NP}$; the second term concerns the relative
velocity $\overrightarrow{v}_{cm,LG}$ between the centers of mass of
the "liquid" complex and the "gas"; the last term represents the
contribution produced by the relative motion of the fragments. A
similar expression can be obtained including others light particles
in the gas "phase" as , for example, the Intermediate Mass Fragments
(IMF).
 From this decomposition we can see how the
isospin equilibration condition ($\overline{\overrightarrow{V}}=0$),
for the total system, requires a very delicate balance which depends
on the average neutron excess ($\beta=\frac{N-Z}{A}$) of the
produced "liquid drops" $\overline{\beta_{L}}$, on the one
associated to the gas "phase" $\overline{\beta_{G}}$, and on the
relative velocities
 between the different parts. To enlighten the role played by some of the terms reported in Eq.(3),
 we can discuss the idealized decay of a
charge/mass asymmetric source through  neutron and proton emission
(or the case in which the liquid drops are produced through a
statistical mechanism $\overline{\overrightarrow{V}_{r,L}}=0$).
Moreover, for simplicity, we can consider uncorrelated fluctuations
between the velocities, masses and neutron excesses. In absence of
pre-equilibrium emission or for identical colliding nuclei
 the second term of
Eq.(3) is zero ($\overline{\overrightarrow{v}_{cm,LG}}= 0$), and the
isospin equilibration requires a neutron or proton gas "phase" or
absence of relative neutron-proton motion. For non- identical
colliding nuclei, if pre-equilibrium emission exists, then
$\overline{\overrightarrow{v}_{cm,LG}}\neq 0$. In this case, if
$\overline{\beta_{G}}\neq \overline{\beta_{L}}$, due, for example,
to the isospin "distillation" phenomenon, the first term has to be
necessarily different from zero and it will contribute to the
neutron-proton differential flow (see also Sect.III).
Therefore, according to our description, the understanding of the
isospin equilibration process for the total system requires the gas
"phase" contribution to be taken into account. This term can be
regarded as a kind of "dissipation" with respect to the system
formed by the liquid part. In this work, as an example, we will
discuss the results obtained through the Constrained Molecular
Dynamics-II approach (CoMD-II) \cite{comd1,comdII} applied to the
charge/mass asymmetric system $^{40}Cl+^{28}Si$ at 40 MeV/nucleon.
The study is performed by using different options for the symmetry
potential term . \vskip 5pt \noindent
 Before to show
the results of our calculations, in the following section we briefly
recall the way in which the isospin interaction  is introduced in
CoMD-II model.

\section{Symmetry interaction and correlations}

According to the results obtained in Ref.\cite{epjnew}, in the
simple case of large and compact systems, the isospin dependent part
of the interaction is expressed, in the so called Non-Local (N.L.),
approximation as:
\begin{eqnarray}
U^{\tau}_{N.L.} & =
&\frac{a_{sym}}{2S_{g.s}}\hat{\rho}A^{2}F'[(1+\frac{1}{2}\alpha
-\alpha')\beta^{2}-\frac{1}{2}\alpha] \\
\alpha'& = & \frac{1}{4}\frac{\partial^{2} \alpha}{\partial
\beta^{2}}|_{\beta=0}
\end{eqnarray}
 $\beta^{4}$ terms are neglected in the previous expression.
$\alpha\equiv\alpha(\hat{\rho})$ represents the correlation
coefficient related to the difference in the dynamics of the n-p
couples with respect to the n-n and p-p ones. It is evaluated at
$\beta=0$ and describes the main effect of the many-body
correlations on the iso-vectorial interaction. $\alpha$ depends on
the average overlap integral per couples of nucleons $\hat{\rho}$
which reflects the degree of compression. The coefficient
$a_{sym}=72 MeV$ determines the strength of the iso-vectorial
interaction at the ground state density ( $g.s.$ ). F' is a form
factor which modulates the changes of the iso-vectorial interaction
as a function of the average overlap integral $S$. In particular for
the Stiff1 option we use $F'=\frac{2S}{S_{g.s.}+S}$, for the Stiff2
case
 $F'=1$ and for the Soft option $F'=(\frac{S_{g.s.}}{S})^{1/2}$.
From eq.(4) we note that the isospin forces if treated in a
self-consistent many-body approach generates, beyond the $\beta^{2}$
dependent potential, also another iso-vectorial density dependent
term, not proportional to $\beta^{2}$, but proportional to the
degree of correlation $\alpha$. As discussed in \cite{epjnew,pyl},
at small asymmetries,  this term determines the high sensitivity of
the experimental observable to the different functional form of
$F'$. The limiting case corresponding to  vanishing values for the
correlation $\alpha$ represents the so called Iso-vectorial Mean
Field Approximation (I.M.F.A.).
 In this case the average overlap integrals per couple of nucleons
related to neutron-neutron, proton-proton and neutron-proton
interactions have the same values and the iso-vectorial interactions
generate only the usual symmetry potential term which is
proportional to $\beta^{2}$ (see eq. 4). A comparison between
results obtained by full CoMD-II calculations and the one derived in
the I.M.F.A. limit will be discussed in the next section.
\section{Calculation results}

 Now we discuss the
results concerning the isospin equilibration process for the
$^{40}Cl+^{28}Si$ system at 40 MeV/nucleon. For this collision, as
an example, in Fig. 1  we show the average total dipolar signals
evaluated through CoMD-II calculations along the $\hat{z}$ beam
direction $\overline{V}^{z}$ and along the impact parameter
direction $\hat{x}$, $\overline{V}^{x}$, respectively. The reference
frame is the c.m. one. The impact parameter $b$ is equal to 3 fm, in
panels (a) and  (c), and 1.5 fm in panels (b) and (d). In Fig.1(a)
and Fig.1(c), the average dipolar signals are shown for the first
150 fm/$c$.
\begin{figure}[htbp]

\includegraphics[width=6cm]{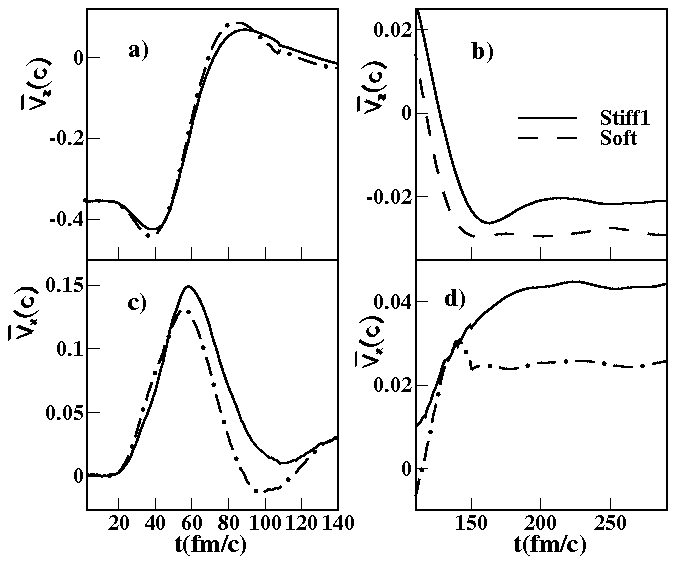}

\caption{Average dipolar signals $\overline{\vec{V}}$ for
b=3 fm along the $\hat{z}$ direction are plotted as a function of
time in the intervals  0-150 fm/c (panel (a)) and 100-400 fm/c
(panel (b)). Different lines indicate different options for the
isospin interaction (see the text). Panels (c) and (d) display the
average dipolar signals  along the $\hat{x}$ direction for b=1.5
fm and in the same time intervals like panels (a) and (b)
respectively.}
\end{figure}
Different lines refer to different iso-vectorial potentials. The
isospin independent compressibility is equal to 220 MeV, according
to Ref. \cite{comd1}. In the first 150 fm/$c$ we can see that in all
the cases wide oscillations exist. They are responsible for the
pre-equilibrium $\gamma$-rays emission \cite{asygdr,ca10mev}. The
damped oscillations converge towards smaller and constant values
(within the uncertainty of the statistics associated to the ensemble
average procedure). This  can be seen in Fig. 1(b) and Fig. 1(d) in
which the dynamical evolution is followed from 100 fm/$c$ up to 300
fm/c. The time interval in which the stationary behavior is reached
is related to the life time of the coherent dipolar collective mode
and it is strictly linked  to the average time for the formation of
the main fragments and pre-equilibrium emission. For the asymptotic
components, global results are shown in Fig.2.
\begin{figure}
\includegraphics[width=6cm]{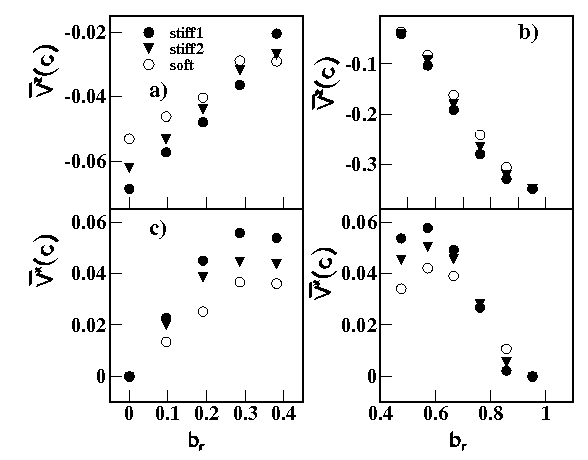}
\caption{Asymptotic values $\overline{V}^{x}$ and $\overline{V}^{z}$ evaluated for
different value of
the $b_{r}$ parameter (see the text). Different symbols indicate different
options for the iso-vectorial interaction.}
\end{figure}
They concern different impact parameters and different interaction
options. In particular, in Fig.3(a), we show for $b$=3 fm the charge
distribution evaluated after 650 fm/c for the Stiff2 option. The
shape of the distribution clearly show that at this impact parameter
the mechanisms evolve essentially throughout a multi-fragmentation
process. Under the same conditions,   we show in Fig.3(b) and   in
Fig.3(c) the ratio $R_{GL}=\frac{|V_{G}^{z}|}{|V_{L}^{z}|}$ between
$\hat{z}$ asymptotic components of the dipolar signal related to the
light particles (first 2 terms of eq.(3)) and the one associated to
the two biggest fragment (last term of eq.(3)). $R_{GL}$ is plotted
as a function of the reduced impact parameter
$b_{r}=\frac{b}{b_{max}}$ ($b_{max}\simeq 7.5 $fm). In Fig.3(b) we
can see that for central and mid-peripheral collisions the "gas"
component (pre-equilibrium contribution) is comparable or larger
than the one associated to the "liquid" one.  Fig.3(c) shows that
the relevance of the "gas" light particle signal rapidly decreases
with the increasing of the impact parameter.

As discussed in the introductory section when the collision partners
are not identical nuclei the isospin equilibration process  produce
a neutron-proton differential flow contribution if the  neutron and
proton "gases" have different c.m. velocities. These pre-conditions
are verified for the studied collision. For $b=3$ fm and for the
Stiff1 and Soft options in Fig.3(d) we show the neutron-proton
differential flow $F_{np}$ as a function of the particles rapidity,
$y$, normalized to the projectile one $y_{beam}$. The rapidity
values are evaluated in the c.m. reference frame. From the figure we
can see that, on average, the neutron-proton transversal velocity
has a negative value. This reflects the  "bending" of the relative
motion between the c.m. of the emitted neutrons and  protons,
through the half-plane opposite to the impact parameter direction.
These results can be compared with the calculations displayed in
Fig.3(e) obtained by subtracting, event by event, the c.m. relative
neutron-proton motion related to the "gas" phase . As can be seen,
similarly to the case  of identical  nuclei, this correction
restores (within the errors associated to the statistics of
simulations) the almost specular of the $F_{np}$ behavior with
respect the rapidity axes. The correction acts also along the beam
direction. The final results also show an enhanced sensitivity to
the options related to the isospin potential.

\begin{figure}
  \includegraphics[width=7cm]{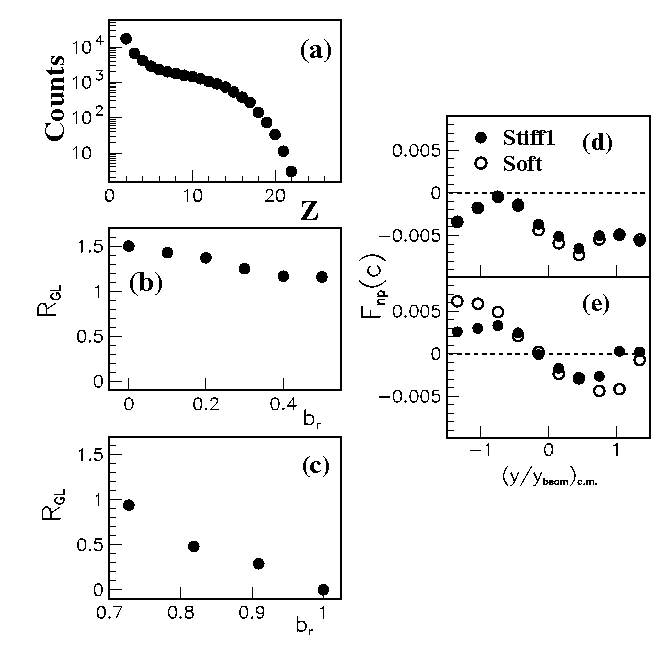}
  \caption{(a) charge $Z$ distribution of the reaction products  obtained
at 650 fm/$c$ for the Stiff2 option and $b$=3 fm. In panels (b) and
(c), the $R_{GL}=\frac{|V_{G}^{z}|}{|V_{L}^{z}|}$ ratio (see the
text) is plotted as a function of the reduced impact parameter.
Panel (d) displays neutron-proton differential flow $F_{np}$ as a
function of the c.m. rapidity. Different symbols refers to different
options describing the iso-vectorial interaction. (e) the same
quantities reported in the panel (d) are plotted after the
correction for the relative velocity associated to the c.m. of the
neutron an protons "gases".}
\end{figure}
In the following we want to discuss in some detail the sensitivity
of the dipolar signal to different options concerning the
iso-vectorial interaction and different approximation scheme. In
Fig.4(a) we  show as a function of the reduced impact parameter
$b_{r}=\frac{b}{b_{max}}$ ($b_{max}\simeq 7.5 $fm) the asymptotic
values of $\overline{V}^{x}$ and $\overline{V}^{z}$. The arrow
indicates the direction of increasing impact parameters
corresponding to the marked points. Different symbols represent
different options. In the region of the bending of the lines, around
$b=3$ fm, we observe the greater sensitivity to the different
options. This result is particularly evident by studying the ratios
$R=\overline{V}^{x}/\overline{V}^{z}$.
\begin{figure}
  \includegraphics[width=6cm]{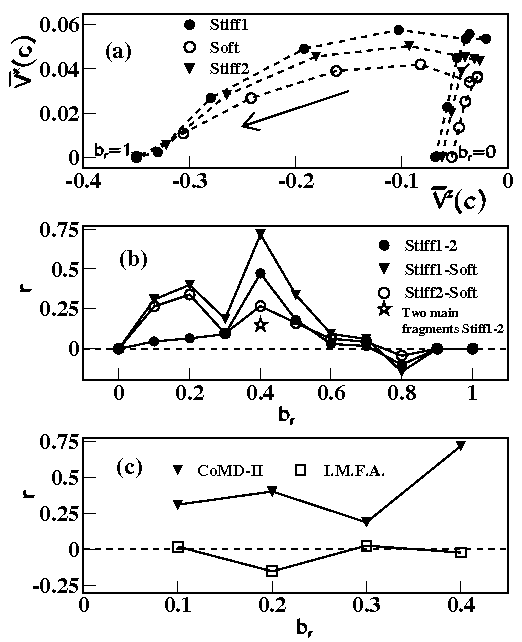}
  \caption{-(a)
 $\overline{V}^{x}$ is plotted as a
 function of the corresponding $\overline{V}^{z}$
 value for different $b_{r}$ values and different options
 (different symbols). The arrow indicates the direction
 of increasing impact parameters.
 -(b) relative changes $r$ for the ratio $R$ (see the text)
 evaluated for different couples of options as a function of
the $b_{r}$ reduced impact parameter. -(c) Values of $r$ evaluated
for the Stiff1-Stiff2 options as a function of $b_{r}$ are plotted
in the case of full CoMD-II calculations and in the case of the
I.M.F.A. approximation (see the text). The lines which join the
points are meant only to guide the eye through the shown trend.}
\end{figure}
In Fig. 4(b) we in fact show the relative change
       $r=\frac{\Delta R}{R}$
 between couples of different options. We can see that for
$b_{r}$ less than about 0.6 large changes are predicted according to
the different form factor shapes. This impact parameter region is
clearly dominated by large overlap between projectile and target
nuclei which gives rise to processes changing from incomplete fusion
reactions to IMF  production. The region of intermediate impact
parameters show the higher sensitivity when the mechanism evolves
with respect to the essentially binary processes which take place at
the higher impact parameters. For $b_{r}$ greater than 0.6, in fact,
the sensitivity is strongly reduced.

In particular, according to what previously observed  (see for
example point (ii)), we have evaluated the partial contributions
$\overline{V}^{x}_{L}$ and $\overline{V}^{z}_{L}$ related to the two
main fragments. As an example, for $b=3$ fm and for the Stiff2
option,  the "liquid" asymptotic values are
$\overline{V}^{x}_{L}=-0.120 c$ and $\overline{V}^{z}_{L}=0.162 c$
while the total contributions are $\overline{V}^{x}=0.044 c$ and
$\overline{V}^{z}=-0.027 c$. It results therefore that the
contributions carried by the two main fragments only partially
contribute to the isospin equilibration process. The remanent part
("gas"), which in this case we have associated to
 particles and to the
IMF, generates a term with opposite sign and similar strength for
both directions. It contributes in a decisive way to the global
equilibration process. In particular, as an example, for $b=3$ fm,
in Fig 4(b) we show with the star symbol, the sensitivity parameter
$r$ evaluated by changing the option from Stiff1 to Stiff2
 and by only taking into account the two main fragments
contributions. As we can see, the partial contribution shows  a
rather reduced sensitivity to the different options as compared to
the case  obtained by using the complete information of the system.

\vskip 5pt \noindent
 Finally, in the following, we show the role of the
correlation coefficient $\alpha$, introduced in Sec.II, into
determine the sensitivity of the investigated observable to the
density dependence of the iso-vectorial interaction. For this aim,
in Fig.4(c), we compare, for different impact parameters, the values
of $r$ obtained for the Stiff2-Stiff1 options with the ones obtained
in the I.M.F.A. limit.
  Fig. 4(c) clearly shows that in
the I.M.F.A. case, at the investigated energies, the sensitivity of
our investigated phenomenon to the behavior of the symmetry
interaction is rather reduced.  The I.M.F.A. also strongly affects
the values of $\overline{\overrightarrow{V}}$. In particular,
independently from the used options, it produces for $b_{r}$=0.4,
values of $|\overline{V}^{z}|$ about four times larger than  the
ones obtained with full COMD-II calculations indicating a reduced
capacity to obtain the isospin equilibration along the $\hat{z}$
direction.

\section{Conclusive remarks}

In summary, in this work the isospin equilibration process for the
asymmetric charge/mass system $^{40}Cl+^{28}Si$ at 40 MeV/nucleon
has been investigated by studying  the ensemble average of the time
derivative of the total dipole $\overline{\overrightarrow{V}}$
evaluated through CoMD-II calculations. Some general properties of
this quantity have been discussed. In particular, it allows to
generalize the definition of isospin equilibration also in complex
reactions evolving through multi-fragmentation processes. CoMD-II
calculations show that the  asymptotic values of
$\overline{\overrightarrow{V}}$ for these processes are quite
sensitive to different symmetry potential options; moreover, in
central and mid-peripheral, the dipolar contribution associated to
the pre-equilibrium emission of charged particles is relevant  to
determine the value of $\overline{\overrightarrow{V}}$ and the
related sensitivity to different density dependent form factors.
CoMD-II calculations performed in the so called I.M.F.A. scheme also
highlights  the fundamental role played by the many-body
correlations in the study of the isospin equilibration processes.


\end{document}